\documentclass[12pt]{article}

\newcommand{\be}{\begin{equation}}
\newcommand{\ee}{\end{equation}}
\newcommand{\bi}[1]{\vspace{-3mm} \bibitem{#1}}
\usepackage{epsfig,amsmath,amssymb,graphics,graphicx}

\begin{document}


\begin{center}
{\Large \bf Conservation Laws and Hamilton's Equations for

\vskip 3mm

Systems with Long-Range Interaction and Memory}
\vskip 7 mm

{\large \bf Vasily E. Tarasov$^{1,2}$, George M. Zaslavsky$^{1,3}$ }\\

\vskip 3mm

{\it $1)$ Courant Institute of Mathematical Sciences, New York University \\
251 Mercer St., New York, NY 10012, USA }\\ 
{\it $2)$ Skobeltsyn Institute of Nuclear Physics, \\
Moscow State University, Moscow 119992, Russia } \\
{\it $3)$ Department of Physics, New York University, \\
2-4 Washington Place, New York, NY 10003, USA } 
\end{center}

\vskip 3mm

\begin{abstract}
Using the fact that extremum of variation of generalized action can lead to
the fractional dynamics in the case of systems with long-range interaction
and long-term memory function, we consider two different applications of 
the action principle: generalized Noether's theorem and Hamiltonian type equations.
In the first case, we derive conservation laws in the form of continuity
equations that consist of fractional time-space derivatives.
Among applications of these results, we consider a chain of 
coupled oscillators with a power-wise memory function and 
power-wise interaction between oscillators.
In the second case, we consider an example of fractional differential 
action 1-form and find the corresponding Hamiltonian type equations from 
the closed condition of the form.
\end{abstract}

\vskip 3mm

\newpage
\section{Introduction}

Different physical phenomena such as anomalous transport or 
random walk with infinite moments \cite{MS,BG},
dynamics of porous media \cite{Nig,GM},
continuous time random walk \cite{A-Z,MK,BMK},
chaotic dynamics \cite{Zaslavsky1} (see also reviews \cite{H,WBG})
can be described by equations with fractional integro-differentiation.
Despite of fairly deep and comprehensive results in fractional calculus
(see \cite{SKM,Podlubny,MR,KST}) 
a possibility of their applications to physics needs to develope 
specific physical tools such as extension of fractional calculus
to the areas as multi-dimension \cite{SKM,MMW},
multi-scaling \cite{MBB,Zm}, 
variational principles \cite{Ag,SB}.
It became recently formalized the links of the dynamical equations with
fractional derivatives to the systems with long-range interaction 
\cite{LZ,TZ3,KZT} and long-term memory \cite{TZ2007}.

In this paper, we concentrate on two problems 
important for numerous physical applications:
conservation laws and Hamiltonian type equations,
both obtained from the corresponding fractional action principles.
In Sec.2, we derive the Noether's theorem for a Lagrangian
that includes non-local space-time densities.
The Noether's theorem was also discussed in \cite{Cresson,FT}. 
Our new derivation shows in an explicit way how fractional derivative 
in time emerges from the specific type of the memory function,
and how fractional derivative in space is related to a specific long-distance
potential of interaction (Sec. 3.)
In Sec. 4, these results are applied to a chain of nonlinear oscillators
that is a subject of great interest in statistics and dynamics \cite{Lieb,Ruffo}.
Finally, at Sec.5, we derive a specific case of fractional Hamilton's equations.
Different steps in this direction were performed in \cite{Riewe,MBal,Stan}.
We consider the Lagrangian density as a functional without fractional derivatives
but, instead, the differential 1-form has fractional differentials.
Some examples are given for this type of systems.

The main feature of this paper is the consideration of fractional type 
differentials or derivatives in both space-time coordinates.

\section{Noether's theorem for long-range interaction and memory}

\subsection{Action and Lagrangian functionals}

Let us consider the action functional
\be  \label{act} S[u] =\int_R d^{2}x \int_R d^{2} y \, 
{\cal L}(u(x),u(y),\partial u(x),\partial u(y)) ,  \ee
where $x=(t,r)$, $t$ is time, $r$ is coordinate, and $y=(t',r')$,
$\partial u(x)=(\partial_t u(t,r),\partial_r u(t,r))$.
The integration is carried out over a region $R$ of the 
$2$-dimensional space $\mathbb{R}^2$ to which $x$ belong.
The field $u(x)$ is defined in the region $R$ of $\mathbb{R}^{2}$.
We assume that $u(x)$ has partial derivatives
\[ \partial_0 u(x)=\frac{\partial u(t,r)}{\partial t}, \quad
\partial_1 u(x)=\frac{\partial u(t,r)}{\partial r},  \]
which are smooth functions with respect to time and coordinate.
Here ${\cal L}(u(x),u(y),\partial u(x),\partial u(y))$ 
is generalized density of Lagrangian. 
If
\be \label{LL} {\cal L}(u(x),u(y),\partial u(x),\partial u(y))=
{\cal L}(u(x),\partial u(x)) \, \delta (x-y) , \ee
then we have the usual action functional
\[ S[u] =\int_R d^{2}x \, {\cal L}(u(x),\partial u(x)) .  \]

The variation of the action (\ref{act}) is
\[ \delta S[u,h]=\int_R d^{2}x \int_R d^{2} y \, 
\Bigl(  \frac{\partial {\cal L} }{\partial u(x)} h(x)+
\frac{\partial {\cal L}}{\partial (\partial_{\mu} u(x))} \partial_{\mu} h(x)+ \]
\be \label{var1}
+\frac{\partial {\cal L} }{\partial u(y)} h(y)+
\frac{\partial {\cal L} }{\partial (\partial_{\mu} u(y))} \partial_{\mu} h(y) 
\Bigr) , \ee
where $\mu=0,1$, $\partial_{\mu}=\partial / \partial x^{\mu}$ and
\[ {\cal L}= {\cal L}(u(x),u(y),\partial u(x),\partial u(y)) , \]
and $h(x)=\delta u(x)$ is the variation of the field $u$. 
The variation (\ref{var1}) can be presented as
\be \label{Suh}
\delta S[u,h]=\int_R d^{2}x \int_R d^{2} y \, 
\left(  \frac{\partial {\cal L}_s}{\partial u(x)} h(x)+
\frac{\partial {\cal L}_s }{\partial (\partial_{\mu} u(x))} \partial_{\mu} h(x)
\right) , \ee
where
\[ {\cal L}_s= 
{\cal L}(u(x),u(y),\partial u(x),\partial u(y))+ {\cal L} (u(y),u(x),\partial u(y),\partial u(x)) . 
\]
We can define the functional
\be \label{LLL}
L[x,u,\partial u] =\frac{1}{2} \int_R d^{2} y \,  {\cal L}_s ,
\ee
which will be called the Lagrangian functional. 
For (\ref{LL}), the functional (\ref{LLL}) is equal to the 
usual Lagrangian density ${\cal L}(u,\partial u)$.

Using notations
\be \label{varL}
\frac{\delta L [x,u,\partial u] }{\delta u(x)}=
\int_R d^{2} y \, \frac{\partial {\cal L}_s}{\partial u(x)} , \quad
\frac{\delta L [x,u,\partial u] }{\delta (\partial_{\mu} u(x))}=
\int_R d^{2} y \, \frac{\partial {\cal L}_s }{\partial (\partial_{\mu} u(x))} ,
\ee
the variation (\ref{Suh}) is
\be \label{5z}
\delta S[u,h]= \int_R d^{2} x \, \left( \frac{\delta L [x,u,\partial u] }{\delta u(x)} h(x)+
\frac{\delta L [x,u,\partial u] }{\delta (\partial_{\mu} u(x))} \partial_{\mu} h(x) \right) . 
\ee
The structure of (\ref{5z}) is non-local, i.e., it includes, through (\ref{LLL})
and ${\cal L}_s$ a possibility of the long-term memory and long-range interaction. 
Let us derive the generalization of the Noether's theorem for this case 
in the way analogical to the local case (\ref{LL}).


\subsection{Equations of motion}

First, we separate the variation that is linked in the variation of coordinates
\be \label{tra}
x^{\mu} \ \rightarrow \ x^{\prime \mu}=x^{\mu}+\delta x^{\mu} ,  
\ee
and the variation caused by a change of the form of $u$,
\[ u(x) \ \rightarrow \  u^{\prime}(x)=u(x)+\delta u(x) . \]
The variation $h(x)=\delta u(x)$ of $u(x)$ that 
is not due to the variation in coordinates is called local.
The total variation is
\be \label{Deltau}
\Delta u(x)= u^{\prime}(x^{\prime})-u(x) =
\delta u(x)+ (\partial u/\partial x^{\mu}) \delta x^{\mu}  
\ee
in the first approximation with respect to $\delta x$. 
Let us consider the variation of (\ref{act}) as
\be \label{av} \delta S[u,h]=
\int_R d^{2} x^{\prime} \, L[x^{\prime} ,u^{\prime} , \partial^{\prime} u^{\prime}] 
- \int_R d^{2} x \, L[x,u,  \partial u] . \ee
The elements of the two-dimensional volume in new and old coordinates are related through the formula
\[ d^{2} x^{\prime} =J(x^{\prime}/x) d^{2} x , \]
where
$J(x^{\prime}/x) =\det \left|  {\partial x^{\prime \mu}}/{\partial x^{\nu}} \right|$
is the Jacobian of the transformation. 
Using the well-known relation
\be 
\det \, A =\exp \ Tr \ln A , 
\ee
and the linear approximation
\[ \frac{\partial x^{\mu \prime}}{\partial x^{\nu}} =
\delta^{\mu}_{\nu}+\partial_{\nu} (\delta x^{\mu}) , \]
we get
\[ J(x^{\prime}/x) =1 +\partial_{\mu} (\delta x^{\mu}) . \]
Here $\delta^{\mu}_{\nu}$ is the Kronecker symbol.

For the variation (\ref{av}), we get
\[ \delta S[u,h]=\int_R d^{2}x \, \left( \delta L+
L  \partial_{\mu} ( \delta x^{\mu}) \right)= \]
\be \label{av28}
=\int_R d^{2}x \, \left( \frac{\delta L}{\delta u} \delta u+ 
\frac{\delta L}{\delta (\partial_{\mu} u)} \delta (\partial_{\mu} u)+  
\partial_{\mu} (L \delta x^{\mu}) \right) , \ee
where the variations of $L$ are defined in (\ref{varL}). 
Using $\delta (\partial_{\mu} u)= \partial_{\mu} (\delta u)$, and
\[ \frac{\delta L}{\delta (\partial_{\mu} u)} \partial_{\mu} \delta u =
\partial_{\mu} \left(
\frac{\delta L}{\delta (\partial_{\mu} u)} \delta u \right) -
\partial_{\mu} \left(
\frac{\delta L}{\delta (\partial_{\mu} u)} \right)  \delta u ,  \]
we can rewrite (\ref{av28}) as
\[ \delta S[u,h]=\int_R d^{2} x \, 
\left( \frac{\delta L}{\delta u}-\partial_{\mu} 
\frac{\delta L}{\delta (\partial_{\mu} u)} \right) \delta u+ \]
\be \label{N10} + \int_R d^{2}x \,
\partial_{\mu} \left(L\delta x^{\mu}+
\frac{\delta L}{\delta (\partial_{\mu} u)} \delta u \right) . \ee
The Gauss theorem gives
\be \label{bi} 
\int_R d^{2}x \, \partial_{\mu} \left(L\delta x^{\mu}+
\frac{\delta L}{\delta (\partial_{\mu} u)} \delta u \right) =
\int_{\partial R} d S_{\mu} \,
\left(L\delta x^{\mu}+\frac{\delta L}{\delta (\partial_{\mu} u)} \delta u \right) . \ee

We assume that at the boundary of the domain of integration the function
$u(x)$ is selected in a definite manner (the boundary condition).
Then stationary (a minimum or a saddle point) values of $S[u]$ from variational 
equations $\delta S[u,h]=0$ with
\[ [\delta u ]_{\partial R}=0  , \quad [\delta x ]_{\partial R}=0 \]
at the boundary $\partial R$ of the domain of integration,
constitutes the necessary and sufficient condition for the real evolution of the field, 
that is, that $u=u(x)$ represents the true dynamics under the given boundary conditions.
The stationary action principle gives
\be \label{ELE} 
\frac{\delta L[x,u,\partial u] }{\delta u(x)}-
\partial_{\mu} \frac{\delta L[x,u,\partial u] }{\delta (\partial_{\mu} u(x))} =0 ,
\ee
where the variations of $L$ are defined in (\ref{varL}). 
This equation is the Euler-Lagrange equation for Lagrangian functional $L[x,u,\partial u]$.

Let us consider three special cases of equation (\ref{ELE}).

(1) The absence of the memory and long-range interaction means that
\[ {\cal L} (u(x),u(y),\partial u(x),\partial u(y)) =
{\cal L} (u(x), \partial u(x)) \, \delta(x-y) . \]
Then (\ref{ELE}) gives the usual Euler-Lagrange equation
\be \frac{\partial {\cal L}(u,\partial u) }{\partial u(x)}-\partial_{\mu} 
\frac{\partial {\cal L}(u, \partial u) }{ \partial  (\partial_{\mu} u(x))} =0 . \ee

(2)  If the generalized Lagrangian density is
\[ {\cal L} (u(x),u(y),\partial u(x),\partial u(y)) =
{\cal L} (u(x),\partial u(x)) \, c_1(D,r) \, \delta(x-y) , \]
where
\[ c_1(D,r)=\frac{|r|^{D-1}}{\Gamma(D)}, \quad (0<D<1) , \]
for a medium distributed on $\mathbb{R}^1$ with the fractional Hausdorff dimension $D$,
then Eq. (\ref{ELE}) has the form
\be
c_1(D,r) \frac{\partial {\cal L} (u,\partial u) }{\partial u(x)}-
\partial­_{\mu} \left( c_1(D,r)
\frac{\partial {\cal L} (u,\partial u) }{\partial (\partial_{\mu} u(x))} \right)=0 .
\ee
This is Euler-Lagrange equation for the field $u(x)=u(t,r)$ in fractal medium.
Examples of the field (wave) equations for fractal medium string and 
fractional hydrodynamics are considered in \cite{Tarasov}. 
For example,
\[ {\cal L}(u(x),\partial u(x))= {1 \over 2} \left( \partial_t u(t,r) \right)^2-
{1 \over 2} v^2 \left( \partial_r u(t,r) \right)^2 , \]
leads to the equation
\[ c_1(D,r) \partial^2_t u(t,r) - v^2
\partial_r \Bigl( c_1(D,r) \partial_r u(t,r) \Bigr)=0   \]
that describes the propagation waves in fractal medium.

(3) Consider the action functional 
\be  \label{ActionN}
S[u]=\int_R d^{2} x \int_R d^{2} y \left( 
\frac{1}{2} \partial_{t} u(x) g_0 (x,y)  \partial_{t'} u(y) 
-\frac{1}{2} \partial_{r} u(x) g_1 (x,y)  \partial_{r'} u(y) 
-V(u(x),u(y)) \right) , \ee
where the kernels $g_0 (x,y)$ and $g_1 (x,y)$ are responsible
for nonlocal time-coordinates dynamics
and $V$ is nonlocal interaction potential. 
Then the Lagrangian functional (\ref{LLL}) is
\[  L[x,u,\partial_t u, \partial_r u] = \]
\be =\int_R d^{2} y \left( 
\frac{1}{2} \partial_{t} u(x) K_0 (x,y)  \partial_{t'} u(y) 
-\frac{1}{2} \partial_{r} u(x) K_1 (x,y)  \partial_{r'} u(y) 
-U(u(x),u(y)) \right) , \ee
where
\[ K_{0}(x,y)=\frac{1}{2}[g_0 (x,y) +g_0 (y,x)] , \] 
\be \label{8A}
K_{1}(x,y)=\frac{1}{2}[g_1 (x,y) +g_1 (y,x)] , \ee
\[ U(u(x),u(y))=V(u(x),u(y))+V(u(y),u(x)) . \]
Assume that 
\[ U(u(x),u(y))=U(u(x)) \delta(x-y). \]
In this case, the Euler-Lagrange functional equation (\ref{ELE}) has the form
(see also \cite{TZ2007})
\be \label{me} 
\int_R d^{2} y \, \partial_{t} K_{0}(x,y) \, \partial_{t'} u(y) -
\int_R d^{2} y \, \partial_{r} K_{1}(x,y) \, \partial_{r'} u(y) +
\frac{\partial U(u(x))}{\partial u(x)} =0 . \ee
It is an integro-differential equation, 
which allows us to derive field equations for different cases of
the kernels $K_{0}(x,y)$ and $K_1(x,y)$.

In the absence of memory and for local interaction the kernels (\ref{8A})
are defined at the only instant $t$ and point $r$, i.e.,
\[ K_{0}(x,y)=g_{0} \delta(x-y) , \quad  K_{1}(x,y)=g_{1} \delta(x-y) \]
with some constants $g_0$ and $g_1$. 
Then equation (\ref{me}) gives 
\[ g_{0} \partial^2_{t} u(t,r) - g_{1} \partial^2_{r} u(t,r) 
+\frac{\partial U(u(t,r))}{\partial u(t,r)} =0 . \]
For example $g_0=g_1=1$, when
\[ U(u(t,r))= -\cos u(t,r) ,  \]
we get the sine-Gordon equation
\be \label{sinG}
\partial^2_t u(t,r)- \partial^2_r u(t,r)+\sin u(t,r)=0  . \ee

\subsection{Noether's current}

Let us derive the Noether's current by 
using the action variation (\ref{N10}).

The second integral of (\ref{N10}) can be presented as
\[ \int_R d^{2}x \,
\partial_{\mu} \left(L\delta x^{\mu}+
\frac{\delta L}{\delta (\partial_{\mu} u)} \delta u \right)=  \]
\be  \label{19z}
=\int_R d^{2} x \, \partial_{\mu} 
\left( \frac{\delta L}{\delta (\partial_{\mu} u)} 
[\delta u+ (\partial_{\nu} u) \delta x^{\nu} ] -
\left[ \frac{\delta L}{\delta (\partial_{\mu} u)} (\partial_{\nu} u) - 
\delta^{\mu}_{\nu} L \right] \delta x^{\nu} \right) . \ee
Using the total variation (\ref{Deltau}), Eq. (\ref{19z}), 
and the energy-momentum tensor
\be \label{emt}
\theta^{\mu}_{\nu} = 
\frac{\delta L}{\delta (\partial_{\mu} u)} (\partial_{\nu} u) - \delta^{\mu}_{\nu} L ,  
\ee
the variation (\ref{N10}) yields 
\[ \delta S[u,h]=\int_R d^{2} x \, 
\left( \frac{\delta L}{\delta u}- \partial_{\mu} 
\frac{\delta L}{\delta (\partial_{\mu} u)} \right) \delta u+ \]
\be \label{action1} 
+ \int_R d^{2} x \,
\partial_{\mu}  \left(
\frac{\delta L}{\delta (\partial_{\mu} u)} \Delta u -
\theta^{\mu}_{\nu} \delta x^{\nu} \right) . \ee

Let us consider a continuous (topological) group of coordinate
transformation $x \ \rightarrow \ x^{\prime}=x^{\prime}(x,a)$,
and let the field function $u=u(x)$ admits the representation
of this group:
\be \label{tr}
x \ \rightarrow \ x^{\prime}, \quad  u(x) \ \rightarrow \ u^{\prime}(x^{\prime}).  \ee
The invariance of functional $S[u]$ with respect to (\ref{tr}) means that
\[ \int_R d^{2} x^{\prime} \, L[x^{\prime} ,u^{\prime} , \partial u^{\prime} /\partial x^{\prime} ] 
= \int_R d^{2} x \, L[x,u , \partial u/\partial x] . \]
The transformation (\ref{tr}) constitutes a group.
Therefore, infinitesimal forms of transformations (\ref{tr}) are
\be 
\Delta x^{\nu}=X^{\nu}_{s} \delta a^{s}  ,  \quad
\Delta u=Y_{s} \delta a^{s} , 
\quad (\nu=0,1, \quad s=1,..,m ) ,
\ee
where $X^{\nu}_{s}$ and $Y_{s}$ are the generators of 
the group of transformation correspondingly in the 
coordinate and the field representations. 
The index $s=1,..,m$ is defined by the representation of the group.

For simple examples of transformation of the coordinate, time or scalar field, we have
\be \label{tr2a}
\Delta x^{\nu}=X^{\nu} \delta a  , \quad (\nu=0,1) ,
\ee
\be \label{tr2b}
\Delta u=Y \delta a . 
\ee

Noether's theorem states that every continuous transformation of coordinate 
(\ref{tr2a}) and field function (\ref{tr2b}), 
which ensures that the variation of the action is zero
admits a conservation law in the form of a continuity equation \cite{Noether,BS}.
Substitution of (\ref{tr2a}), (\ref{tr2b}), and (\ref{ELE}) into (\ref{action1}) gives
\be \label{e31} \delta S[u,h] = \int_R d^{2} x \, 
\partial_{\mu} \left(
\frac{\delta L}{\delta (\partial_{\mu} u)} Y -
\theta^{\mu}_{\nu} X^{\nu} \right) \delta a . \ee
In view of the fact that the variation of the parameter, $\delta a$,
is arbitrary, from (\ref{e31}) we get the conservation law
\be \label{con} 
\partial_{\mu} J^{\mu} =0,   \ee
where
\be \label{Jmk}
J^{\mu} =\frac{\delta L}{\delta (\partial_{\mu} u)} Y -
\theta^{\mu}_{\nu} X^{\nu} ,
\ee
is the Noether's currents and $\theta^{\mu}_{\nu}$ is defined in (\ref{emt}).
Equation (\ref{con}) means that there exists conservation law.
Nontriviality of equations (\ref{con}) and (\ref{Jmk}) is that $L$ 
and $\theta^{\mu}_{\nu}$ have nonlocal interaction and memory.

\section{Application of the Noether's theorem for 
long-range interaction and long-term memory}

\subsection{Lagrangian functional and energy-momentum tensor}

For $\mu=0,1$, $x^0=t$, and $x^1=r$, 
the Euler-Lagrange functional equation (\ref{ELE}) is
\be \label{eqmot}
\frac{\delta L[x,u,\partial_t u, \partial_r u] }{\delta u(x)}-\frac{\partial}{\partial t} 
\frac{\delta L[x,u,\partial_t u, \partial_r u] }{\delta (\partial_{t} u(x))}
-\frac{\partial}{\partial r} 
\frac{\delta L[x,u,\partial u] }{\delta (\partial_{r} u(x))}  =0 . \ee
The time and space variables in action can be separated 
to consider the field with power-law memory and long-range interaction.
Let $K_0(x,y)$ and $K_1(x,y)$ have the form 
\be \label{MM1} K_0 (x,y)=\delta(r-r') {\cal K}_0 (t,t') , \ee
\be \label{MM2} K_1 (x,y)=\delta(t-t') {\cal K}_1(r,r') , \ee
and
\be U(u(x),u(y))=U(u(x))\delta(x-y) , \ee
where $x=(t,r)$, and $y=(t',r')$.  
Then the Lagrangian functional is
\be
L[x,u,\partial_t u, \partial_r u] 
= \frac{1}{2} \int dt' {\cal K}_0(t,t') \partial_{t} u \partial_{t'} u 
-\frac{1}{2} \int dr' {\cal K}_1(r,r') \partial_{r} u \partial_{r'} u -U(u) .
\ee
To present the long-term memory and long-range interaction,
consider the kernels ${\cal K}_{0} (t,t')$ and ${\cal K}_{1} (r,r')$ 
in the power-law forms
\be \label{Mrr} 
{\cal K}_1 (r,r') = {\cal K}_1 (|r-r'|)=
\frac{-g}{\cos (\pi \alpha /2) \Gamma(2-\alpha) }
\frac{1}{|r-r'|^{\alpha-1}} , \quad (1<\alpha<2) ,
\ee
and
\be \label{Mtt} 
{\cal K}_{0} (t,t')=
\begin{cases} 
{\cal M} (t-t') , & 0 < t' < t ;
\cr 
0, & t' > t, \quad \ t'< 0 ,
\end{cases}
\ee
with
\be \label{Mt1}
{\cal M}(t-t') = \frac{1}{\Gamma(1-\beta)} \frac{1}{(t-t')^{\beta}} , 
\quad (0<\beta <1) ,
\ee
Then the variational derivatives of Lagrangian functional are
\be \label{d1}
\frac{\delta L}{\delta u} = - \frac{\partial U(u(t,r))}{\partial u(t,r)} ,
\ee
\be \label{d2}
\frac{\delta L}{\delta (\partial_{t} u)} = \ _{0}^CD^{\beta}_{t} u(t,r) ,
\quad (0<\beta<1) ,
\ee
\be \label{d3}
\frac{\delta L}{\delta (\partial_{r} u)} = 
-g \frac{\partial^{\alpha-1}}{\partial |r|^{\alpha-1}} u(t,r) ,
\quad (1<\alpha<2) ,
\ee
where $ \ _{0}^CD^{\beta}_{t} $ is Caputo derivative \cite{KST,Podlubny} defined by
\[
_0^CD^{\beta}_{t}u(t,r)=
\frac{1}{\Gamma(n-\beta)} 
\int^{t}_{0} \frac{d \tau}{(t-\tau)^{\beta-n+1}} \frac{\partial^n u(\tau,r)}{\partial \tau^n} , 
\quad (n-1<\beta<n).
\]
Substitution of (\ref{d1}), (\ref{d2}) and (\ref{d3}) into (\ref{eqmot}) gives
the fractional field Euler-Lagrange equation
\be \label{main1}
\ _{0}^CD^{\beta+1}_{t} u(t,r) -
g \frac{\partial^{\alpha}}{\partial |r|^{\alpha}} u(t,r)+
\frac{\partial U(u(t,r))}{\partial u(t,r)} =0 ,
\quad (1<\alpha<2, \ 0<\beta<1) .
\ee
For $\alpha=2$ and $\beta=1$, Eq. (\ref{main1}) is the usual field equation
\be \label{usual}
\partial^2_t u(t,r) - g \partial^2_r u(t,r)+
\frac{\partial U(u(t,r))}{\partial u(t,r)} =0 .
\ee
For the potential $U(u(t,r))= -\cos u(t,r)$, 
equation (\ref{usual}) gives the sine-Gordon equation, 
\be 
\partial^2_t u(t,r)- \partial^2_r u(t,r)+\sin u(t,r)=0  , \ee
and Eq. (\ref{main1}) is a spatio-temporal fractional sine-Gordon equation 
\be
\partial^{\beta}_tu-\partial^{\alpha}_{|r|}u+\sin\, u=0 ,
\ee
where we used for abreviation simplified notation 
for fractional derivatives. 
For the case $\beta=2$, the equation was obtained in \cite{LZ}.


The energy-momentum tensor $\theta^{\mu}_{\nu}$, 
can be presented in the form
\be \label{ttt}
\theta^{\mu}_{\nu} = \theta \delta_{\mu \nu} + \tau^{\mu}_{\nu} ,
\ee
where the first term is the diagonal part
\be
\theta=\theta^0_0+\theta^1_1=
\frac{\delta L}{\delta (\partial_t u)} (\partial_t u) +
\frac{\delta L}{\delta (\partial_r u)} (\partial_r u) - 2 L =U(u(t,r)) .
\ee
This  represents a pressure-like quantity.
The second term in the rigt hand side of (\ref{ttt}) 
is a nondiagonal  of $\theta^{\mu}_{\nu}$:
\be
\tau^{\mu}_{\nu}=
\frac{\delta L}{\delta (\partial_{\mu} u)} (\partial_{\nu} u) - 
\frac{\delta L}{\delta (\partial_{\kappa} u)} (\partial_{\kappa} u) \, \delta^{\mu}_{\nu} , 
\ee
where
\be
\tau^0_0=\tau^1_1=0 ,
\ee
\be
\tau^{0}_{1}=\frac{\delta L}{\delta (\partial_t u)} (\partial_r u) =
\Bigl[\ _{0}^CD^{\beta}_{t} u(t,r) \Bigr] \, \partial_r u(t,r) ,
\ee
\be
\tau^1_0=\frac{\delta L}{\delta (\partial_r u)} (\partial_t u) =
-g \Bigl[ \frac{\partial^{\alpha-1}}{\partial |r|^{\alpha-1}} u(t,r) \Bigr] \, \partial_t u(t,r) .
\ee
For $r \in \mathbb{R}^3$ the spatial components of $\tau^{\mu}_{\nu}$, 
represent shear stress tensor. 
The value $\theta$ represents a pressure-like quantity, normal stress.


\subsection{Homogeneity in time}

The homogeneity in time means invariance of action with
respect to the transformation
\be \label{48}
t \rightarrow t^{\prime}=t+a , \quad r \rightarrow r, \quad u \rightarrow u .
\ee
Then the infinitesimal transformations are
\be
\Delta x^{\mu}=\delta^{\mu}_{0} \delta a, \quad \Delta u=0 , \quad (\mu=0,1)
\ee
with the generators 
\be
X^{\mu}=\delta^{\mu}_{0}, \quad Y=0 , \quad (\mu=0,1) .
\ee
The Noether's current has two components 
\be 
J^{\mu} =-\theta^{\mu}_{0} = -\left(
\frac{\delta L}{\delta (\partial_{\mu} u)} \partial_t u - \delta^{\mu}_{0} L  \right),
\quad (\mu=0,1).
\ee

Using $x^0=t$, and $x^1=r$, we get the continuity equation 
\be
\partial_t J^0+ \partial_r J^1=0 ,
\ee
where
\[ 
J^0 =-\left( \frac{\delta L}{\delta (\partial_{t} u)} (\partial_{t} u) - L  \right)= 
- \left( \partial_{t} u(t,r) \ _{0}^CD^{\beta}_{t} u(t,r) - L \right)=
\]
\be 
=-\left( \frac{1}{2} \partial_{t} u(t,r) \ _{0}^CD^{\beta}_{t} u(t,r) +
\frac{1}{2} g \partial_{r} u(t,r)  
\frac{\partial^{\alpha-1}}{\partial |r|^{\alpha-1}} u(t,r) +U \right) ,
\ee
and
\be 
J^1 =-\frac{\delta L}{\delta (\partial_{r} u)} (\partial_{t} u) =
g \partial_{t} u(t,r) \frac{\partial^{\alpha-1}}{\partial |r|^{\alpha-1}} u(t,r) . 
\ee
As a result, the continuity equation is
\be \label{first}
\partial_t \left(  \frac{1}{2} \partial_{t} u \ _{0}^CD^{\beta}_{t} u +
g \frac{1}{2} \partial_{r} u \frac{\partial^{\alpha-1}}{\partial |r|^{\alpha-1}} u  +U \right)
- \partial_r 
\left( g \partial_{t} u \frac{\partial^{\alpha-1}}{\partial |r|^{\alpha-1}} u \right)=0  ,
\ee
where $u=u(t,r)$. Here 
\be \label{Hd1}
{\cal H}= \frac{1}{2} \partial_{t} u \ _{0}^CD^{\beta}_{t} u +
\frac{1}{2} g \partial_{r} u \frac{\partial^{\alpha-1}}{\partial |r|^{\alpha-1}} u  +U 
\ee
is a fractional generalization of density of energy (density of Hamiltonian), and
\be \label{PPP}
{\cal P}=-g \partial_{t} u \frac{\partial^{\alpha-1}}{\partial |r|^{\alpha-1}} u 
\ee
is a fractional generalization of density of momentum.
For the case $\alpha=2$ and $\beta=1$, we obtain the well-known relations \cite{BS}:
\be
{\cal H}= \frac{1}{2} (\partial_{t} u)^2 + \frac{1}{2} g (\partial_{r} u)^2 +U ,
\quad {\cal P}=-g \partial_{t} u \partial_r u .
\ee
The continuity equation (\ref{first}) 
can be presented in an usual form 
of the conservation of energy
\be \label{PH}
\partial_t {\cal H}+\partial_r {\cal P}=0 
\ee
with the fractional generalizations of the energy and momentum 
given by (\ref{Hd1}) and (\ref{PPP}).

\subsection{Homogeneity of space}

The homogeneity of space means invariance of action with
respect to transformations
\be
r \rightarrow r^{\prime}=r+a , \quad t \rightarrow t ,  \quad u \rightarrow u .
\ee
The corresponding infinitesimal transformations are presented by
\be
\Delta x^{\mu}=\delta^{\mu}_{1} \delta a, \quad \Delta u=0 
\ee
with the generators 
\be
X^{\mu}=\delta^{\mu}_{1}, \quad Y=0.
\ee
Then the Noether's current has the following two components
\be 
J^{\mu} =-\theta^{\mu}_{1} = -\left(
\frac{\delta L}{\delta (\partial_{\mu} u)} (\partial_r u) - \delta^{\mu}_1 L  \right),
\quad (\mu=0,1).
\ee

Using $x^0=t$, and $x^1=r$, we get the continuity equation
that corresponds to the homogeneity of one-dimensional space
\be \label{ce2}
\partial_t J^0+ \partial_r J^1=0 ,
\ee
where
\be 
J^{0} =
-\frac{\delta L}{\delta (\partial_{t} u)} (\partial_{r} u)  =
- \partial_{r} u(t,r)  \ _{0}^CD^{\beta}_{t} u(t,r) ,
\ee
and
\[ 
J^{1} =-\frac{\delta L}{\delta (\partial_{r} u)} (\partial_{r} u) + L =
g \partial_{r} u(t,r) \frac{\partial^{\alpha-1}}{\partial |r|^{\alpha-1}} u(t,r) +L =
\]
\be 
= \frac{1}{2} \partial_{t} u(t,r)  \ _{0}^CD^{\beta}_{t} u(t,r) 
+\frac{1}{2}g \partial_{r} u(t,r) 
\frac{\partial^{\alpha-1}}{\partial |r|^{\alpha-1}} u(t,r) -U .
\ee
As a result, the continuity equation (\ref{ce2}) is
\be \label{second}
-\partial_t \left(  \partial_{r} u  \ _{0}^CD^{\beta}_{t} u \right)+
\partial_r \left(  \frac{1}{2} \partial_{t} u  \ _{0}^CD^{\beta}_{t} u 
+\frac{1}{2} g \partial_{r} u \frac{\partial^{\alpha-1}}{\partial |r|^{\alpha-1}} u -U
\right)=0 
\ee
that can be interpreted as the momentum conservation law
in the case of fractional dynamics.

For $\beta=1$ and $\alpha=2$, equation (\ref{second}) gives
\be
-\partial_t \left(  \partial_{r} u  \partial_t u \right)+
\partial_r \left(  \frac{1}{2} (\partial_{t} u)^2 
+\frac{1}{2} g (\partial_{r} u )^2 -U \right)=0 .
\ee
Note that, 
in general, for integer and fractional values of $\alpha$  and $\beta$,
$J^0_1 \not= J^1_0$, and the energy-momentum tensor
$\theta^{\mu}_{\nu}$ is not a symmetric with respect to $\mu$ and $\nu$. \\

\subsection{Field translation invariance}

For the case $U(u)=0$, the action (\ref{ActionN}) is invariant 
with respect to transformations
\be
x^{\mu} \rightarrow x^{\prime \mu}=x^{\mu} , \quad u \rightarrow u +a .
\ee
The infinitesimal transformations are presented as
\be
\Delta x^{\mu}=0 , \quad \Delta u= \delta a 
\ee
with the generators 
\be
X^{\mu}=0 , \quad Y=1.
\ee
The Noether's current 
\be 
J^{\mu} = \frac{\delta L}{\delta (\partial_{\mu} u)} , \quad (\mu=1,2)
\ee
has components
\be
J^0=\frac{\delta L}{\delta (\partial_{t} u)} =\ _{0}^CD^{\beta}_{t} u(t,r) ,
\quad (0<\beta<1) ,
\ee
\be
J^1=\frac{\delta L}{\delta (\partial_{r} u)} = 
-g \frac{\partial^{\alpha-1}}{\partial |r|^{\alpha-1}} u(t,r) ,
\quad (1<\alpha<2) .
\ee
As a result, the continuity equation
\[ \partial_{\mu}J^{\mu}=\partial_t J^0+\partial_rJ^1=0 \]
has the form
\be \label{3law}
\partial_t \ _{0}^CD^{\beta}_{t} u(t,r) 
-g  \partial_r \frac{\partial^{\alpha-1}}{\partial |r|^{\alpha-1}} u(t,r)=0 ,
\quad (1<\alpha<2, \ 0<\beta<1) .
\ee
For $\alpha=2$ and $\beta=1$, we get
\be 
\partial^2_t u(t,r) -g  \partial^2_r u(t,r)=0 ,
\ee
which is the usual field equation (\ref{usual}) for $U(u)=0$.

Note that \cite{KST}
\be 
\partial_t \ _{0}^CD^{\beta}_{t} u(t,r) = \ _{0}^CD^{1+\beta}_{t} u(t,r) 
+\frac{t^{-\beta}}{\Gamma(1-\beta)} \partial_t u(0,r) . 
\ee
To have the relation 
$\partial_t \ _{0}^CD^{\beta}_{t} u(t,r) = \ _{0}^CD^{1+\beta}_{t} u(t,r)$,
the initial conditions $\partial_t u(0,r) =0$ should be applied. 
In general, equation (\ref{3law}) cannot be presented as
\be
\ _{0}^CD^{1+\beta}_{t} u(t,r) 
-g  \frac{\partial^{\alpha}}{\partial |r|^{\alpha}} u(t,r)=0 ,
\quad (1<\alpha<2, \ 0<\beta<1) ,
\ee
which is the fractional field equation (\ref{main1}) for $U(u)=0$.

This shows that the conservation law (\ref{3law}), in general,
doesn't coincide with the field equation as it is happened for
integer derivatives, unless, we use special boundary and 
initial conditions.

\section{Chain with long-range interaction}

It was shown in \cite{LZ,TZ3,KZT,KZ,JMP} how the long-range interaction 
between different oscillators can be described by the fractional differential equations 
in the continuous medium limit. 
In this section, the Noether's theorem will be applied to such kind of systems.

\subsection{Equation of motion and Noether's currents}

Let us define the action as
\be \label{Sun}
S[u_n]=\int^{+\infty}_{-\infty} dt \Bigl(
\sum_{n=-\infty}^{+\infty}
\left[ \frac{1}{2} \; \dot{u}^2_n(t)  -V (u_n(t)) \right] - 
\sum_{\substack{n, m=-\infty \\ m \ne n}}^{+\infty} \;  U (u_n(t),u_m(t)) \Bigr) ,
\ee
where $u_n$ is displacement of the $n$-th oscillator from the equilibrium, 
\be \label{UU}
U(u_n(t))= \frac{1}{4} g_0 J_{\alpha}(|n-m|) \, (u_n(t)-u_m(t))^2  ,
\ee
and
\be \label{189}
J_{\alpha}(|n-m|) =\frac{1}{|n-m|^{\alpha+1}}, \quad (\alpha>0) .
\ee
The Lagrangian of the chain is
\be \label{chainL}
{\cal L}=\sum_{n=-\infty}^{+\infty}
\left[ \frac{1}{2} \; \dot{u}^2_n(t) - V (u_n(t)) \right] - 
\frac{1}{4} g_0 \sum_{\substack{n, m=-\infty \\ m \ne n}}^{+\infty} \;  
J_{\alpha}(|n-m|) \, (u_n(t)-u_m(t))^2  . 
\ee
The equation of motion 
\be
\frac{\partial {\cal L} }{\partial u_n(t)}-
\frac{d}{dt} \frac{\partial {\cal L} }{\partial \dot{u}_n(t)}=0
\ee
for Lagrangian (\ref{chainL}) has the form
\be \label{cem}
\frac{d^2}{dt^2} u_n(t) +\frac{\partial V(u_n)}{\partial u_n(t)} +
g_0 \sum_{\substack{n, m=-\infty \\ m \ne n}}^{+\infty} \;  
J_{\alpha}(|n-m|) \, [u_m(t)-u_n(t)] =0 .
\ee
A continuous limit of equation (\ref{cem}) can be defined by
a transform operation from $u_n(t)$ to $u(x,t)$
\cite{LZ,TZ3,KZT,KZ,JMP}. 
First, define $u_n(t)$ as Fourier coefficients of some 
function $\hat{u}(k,t)$, $k \in [-K/2, K/2]$, i.e.
\be \label{ukt}
\hat{u}(t,k) = \sum_{n=-\infty}^{+\infty} \; u_n(t) \; e^{-i k x_n} =
{\cal F}_{\Delta} \{u_n(t)\} ,
\ee
where  $x_n = n \Delta x$, and $\Delta x=2\pi/K$ is a
distance between nearest particles in the chain, and
\be \label{un} 
u_n(t) = \frac{1}{K} \int_{-K/2}^{+K/2} dk \ \hat{u}(t,k) \; e^{i k x_n}= 
{\cal F}^{-1}_{\Delta} \{ \hat{u}(t,k) \} . 
\ee
Secondly, in the limit $\Delta x \rightarrow 0$ ($K \rightarrow \infty$) 
replace $u_n(t)=(2\pi/K) u(x_n,t) \rightarrow  u(x,t) dx$, 
and  $x_n=n\Delta x= 2\pi n/K \rightarrow x$.
In this limit, Eqs. (\ref{ukt}), (\ref{un}) are transformed into 
the integrals
\be \label{ukt2} 
\tilde{u}(t,k)=\int^{+\infty}_{-\infty} dx \ e^{-ikx} u(t,x) = 
{\cal F} \{ u(t,x) \} = \lim_{\Delta x \rightarrow 0} {\cal F}_{\Delta} \{u_n(t)\} , 
\ee
\be \label{uxt}
u(t,x)=\frac{1}{2\pi} \int^{+\infty}_{-\infty} dk \ e^{ikx} \tilde{u}(t,k) =
 {\cal F}^{-1} \{ \tilde{u}(t,k) \}= 
\lim_{\Delta x \rightarrow 0} {\cal F}^{-1}_{\Delta} \{ \hat{u}(t,k) \}  . 
\ee

Applying (\ref{ukt}) to (\ref{cem})
and performing the limit (\ref{ukt2}), we obtain
\be \label{200}
\frac{\partial^2 u(t,x)}{\partial t^2}+
g_{\alpha} \frac{\partial^{\alpha} u(t,x)}{\partial |x|^{\alpha}}+
\frac{\partial V(u)}{\partial u(t,r)}=0, \quad (0<\beta<2, \ 1<\alpha<2) ,
\ee
where 
\be
g_{\alpha}=2 g_0 (\Delta x)^{\alpha}  \Gamma(-\alpha) \cos 
\left( \frac{\pi \alpha}{2} \right)
\ee
is the renormalized constant. 
Equations (\ref{200}) were considered in \cite{LZ,TZ3,KZT,KZ,JMP}.

Consider a continuous transformation of time and field 
\be \label{tr3}
t \ \rightarrow \ t^{\prime}, \quad  u_n(t) \ \rightarrow \ u^{\prime}_n(t^{\prime}),  \ee
which form a continuous group with generators $X$ and $Y_n$ such that
the infinitesimal transformations are
\be 
\Delta t=X \delta a  , \quad  \Delta u_n=Y_{n} \delta a . 
\ee
The corresponding Noether's current is
\be 
J = \sum_{n=-\infty}^{+\infty}  
\frac{\partial {\cal L} }{\partial \dot{u}_n(t) } Y_{n} - \theta X ,
\ee
where 
\[ \theta = \sum_{n=-\infty}^{+\infty}
\frac{\partial {\cal L} }{\partial \dot{u}_n(t) } \dot{u}_n(t)  - {\cal L}  \]
is the energy. Then
\be 
J = \sum_{n=-\infty}^{+\infty}  
\frac{\partial {\cal L} }{\partial \dot{u}_n(t) } 
\Bigl[ Y_{n} - X \dot{u}_n(t) \Bigr] + {\cal L} X.  
\ee
The equation of conservation law is
\be 
\frac{d}{dt} J =0  .  
\ee
In the next subsections, we consider examples of the conservation laws. 

\subsection{Homogeneity of time}

The homogeneity of time means the invariance of action 
with respect to the transformation (compare to (\ref{48}))
\be
t \rightarrow t+a, \quad u_n \rightarrow u_n .
\ee
Its infinitesimal form is
\be
\Delta t= \delta a , \quad \Delta u_n=0 ,
\ee
with generators
\be
X=1, \quad Y_{n}=0 .
\ee
Then the Noether's current is
\[
J_t = -\theta =- \sum_{n=-\infty}^{+\infty}
\frac{\partial {\cal L} }{\partial \dot{u}_n(t) } \dot{u}_n(t) + {\cal L} =  
- \sum_{n=-\infty}^{+\infty} \dot{u}^2_n(t)  + {\cal L}  =  
\]
\[
=- \sum_{n=-\infty}^{+\infty}
 \frac{1}{2} \; \dot{u}^2_n(t) -
\sum_{n=-\infty}^{+\infty} V (u_n(t)) -
\frac{1}{4} g_0\sum_{\substack{n, m=-\infty \\ m \ne n}}^{+\infty} \;  
J_{\alpha}(|n-m|) \, (u_n(t)-u_m(t))^2  \equiv -H , 
\]
where $H$ is the Hamiltonian.
The conservation law $dH/dt=0$ is
in the continuous limit gives the equation
\be \label{chain-c1}
\frac{d}{dt} \int^{\infty}_{-\infty} dr \, 
\left[ \frac{1}{2} \; [\partial_t u(t,r)]^2 + V (u(t,r)) +
\frac{1}{2} g_0 , u(t,r)  
\frac{\partial^{\alpha}}{\partial |r|^{\alpha}} u(t,r) \right]=
\frac{d}{dt} \int^{+\infty}_{-\infty} dr \, {\cal H}=0 , \ee
where $1<\alpha<2$, and the density ${\cal H}$ of Hamiltonian (\ref{Hd1})
is introduced. 

One can compare this equation with equation (\ref{PH}) for $\beta=1$.
Both results coincide if the boundary conditions
\be
\lim_{r \rightarrow \pm \infty} {\cal P}=0 
\ee
is applied.

\subsection{Translation invariance}

For $V(u_n)=0$, the action (\ref{Sun}) is invariant 
with respect to the transformations
\be
t \rightarrow t, \quad u_n \rightarrow u_n +a ,
\ee
or in the infinitesimal form 
\be
\Delta t= 0 , \quad \Delta u_n= \delta a ,
\ee
with the generators 
\be
X=0, \quad Y_{n}=1 .
\ee
Then the Noether's current is 
\be
J_r = \sum_{n=-\infty}^{+\infty}  \frac{\partial {\cal L} }{\partial \dot{u}_n(t) } 
=\sum_{n=-\infty}^{+\infty} \; \dot{u}_n(t) \equiv P ,
\ee
where $P$ is the tolal momentum.
The conservation law is $dP/dt=0$.

The continuous limit of this conservation law gives the equation
\be \label{dPdt}
\frac{d}{dt} \int^{+\infty}_{-\infty} dr \, \partial_t u(t,r) =0 . 
\ee

Let us compare this equation with Eq. (\ref{3law})
that is derived for scalar field $u(t,r)$.
Integration of (\ref{3law}) with respect to coordinate $r$ gives 
\be 
\partial_t \int^{+\infty}_{-\infty} dr \ _{0}^CD^{\beta}_{t} u(t,r) 
-g  \int^{+\infty}_{-\infty} dr  \partial_r 
\frac{\partial^{\alpha-1}}{\partial |r|^{\alpha-1}} u(t,r)=0 ,
\quad (1<\alpha<2, \ 0<\beta<1) .
\ee
Then
\be 
\partial_t \int^{+\infty}_{-\infty} dr \ _{0}^CD^{\beta}_{t} u(t,r) 
-g  \Bigl(
\frac{\partial^{\alpha-1}}{\partial |r|^{\alpha-1}} u(t,r) \Bigr)^{+\infty}_{-\infty}=0 .
\ee
As a result,  equation (\ref{3law}) coincides with (\ref{dPdt}), 
for $\beta=1$, if we use the boundary conditions
\be
\lim_{r \rightarrow \pm \infty} 
\frac{\partial^{\alpha-1}}{\partial |r|^{\alpha-1}} u(t,r)=0 .
\ee


\section{Fractional Hamilton's equations}

\subsection{De Donder-Weyl Hamiltonian}

The idea of variation of fields based on a manifestly 
covariant version of the Hamiltonian formalism in field theory 
known in the calculus of variation of multiple integrals \cite{Dedonder,Kastrup} 
has been proposed by M. Born and H. Weyl \cite{BW}. 
The mathematical study of 
geometrical structures underlying the related aspects of 
the calculus of variations and classical field theory 
has  been undertaken recently by many authors 
(see for example \cite{Kastrup,Sardan,Kanat}).  

The De Donder-Weyl Hamiltonian form of the field equations \cite{Dedonder} are
\be \label{155}
\partial_{\mu} u (x) = \frac{\partial {\cal H}}{\partial \pi^{\mu}},  
\quad 
\partial_\mu \pi^{\mu} (x) = - \frac{\partial {\cal H}}{\partial u} ,  
\ee 
where 
\be \label{e37} 
\pi^{\mu} = \frac{\partial {\cal L}}{\partial (\partial_{\mu} u)} , \ee
is called the multi-momenta, and 
\be \label{calH}
{\cal H} ( u, \pi^{\mu}) = (\partial_{\mu} u)\, \pi^{\mu} - {\cal L}  
\ee 
is called the De Donder-Weyl Hamiltonian density function,  
${\cal L} = {\cal L}(u, \partial_{\mu} u)$ is Lagrangian density.  
These equations are known to be equivalent 
to the Euler-Lagrange field equations if 
${\cal L}$ is regular in the sense that 
\be  
\det \left[ \frac{\partial^2 {\cal L}}{
\partial (\partial_{\mu} u ) \partial (\partial_{\mu} u) } \right] \neq 0. 
\ee


\subsection{Hamilton's equations of integer order}

Let us consider Hamiltonian systems in the extended phase 
space of coordinates $(x^{\mu},u,\pi^{\mu})$.
The evolution of fields is defined by 
stationary states of the action functional
\be \label{Sqp}
S[u,\pi]=\int [\pi^{\mu} \partial_{\mu} u - {\cal H}(u,\pi)]  \, d^2 x,
\ee
where the Hamiltonian density ${\cal H}$ is defined by (\ref{calH}),
both $u$ and $\pi$ are assumed to be independent functions of $x=(t,r)$.
In classical field theory, the evolution of the field $u(x)$ is derived 
by finding the condition for which the action integral (\ref{Sqp}) 
is stationary (a minimum or a saddle point). 
The action functional (\ref{Sqp}) can be rewritten as 
\be \label {S38}
S[u,\pi]=\int \omega_{\mu} ,
\ee
where
\be \label{CP}
\omega_{\mu}=\pi^{\mu} d u - {\cal H} ( u, \pi )  dx^{\mu} .
\ee
is the Poincare-Cartan 1-form or the action 1-form.

The stationary action condition $\delta S=0$ leads to
\be \label{domega}
d \omega_{\mu}= 0 .
\ee
Here the exterior derivative is
\be
d=(dx^{\nu}) D_{x^{\nu}}+(du) D_u+(d \pi^{\nu}) D_{\pi^{\nu}} ,
\ee
where we put new notation for the derivatives
\be 
D_{x^{\nu}} =\frac{\partial }{\partial x^{\nu}}, \quad
D_u =\frac{\partial }{\partial u}, \quad
D_{\pi^{\nu}} =\frac{\partial }{\partial {\pi^{\nu}}} . 
\ee

The De Donder-Weyl Hamiltonian form of field equations
in (\ref{domega}) can be obtained from the condition 
that the derivative is exterior one.
This condition is equivalent to the 
stationary action principle $\delta S[u,\pi]=0$. 
Then Eq. (\ref{CP}) gives
\[ d\omega_{\mu}=d(\pi^{\mu} du)-d({\cal H}dx^{\mu})=
(D_{x^{\nu}} \pi^{\mu}) d x^{\nu} \wedge d u+
(D_u \pi^{\mu}) d u \wedge d u+ (D_{\pi^{\nu}} \pi^{\mu}) d \pi^{\nu} \wedge d u - \]
\be \label{dCP1}
-(D_{x^{\nu}} {\cal H}) d x^{\nu} \wedge d x^{\mu}
- (D_u {\cal H}) d u \wedge d x^{\mu}-(D_{\pi^{\nu}} {\cal H}) d \pi^{\nu} \wedge d x^{\mu}.
\ee
Using $d {\pi} \wedge d x=-d x \wedge d {\pi}$, and $(D_{x^{\nu}} {\cal H})$, 
$D_u {\pi}=0$, we get
\be \label{dCP2}
d\omega=\left( D_x {\pi}+D_u {\cal H} \right)  d x \wedge d u-
\left( D_{\pi} {\pi} d u-D_{\pi} {\cal H} d x \right) \wedge d {\pi} ,
\ee
where the matrix form of the equation is used. 
The relation $D_{\pi} {\pi}=1$ gives
\be \label{dCP3e}
d\omega=\left( D_x {\pi}+D_u {\cal H} \right)  d x \wedge d u-
\left( d u-D_{\pi} {\cal H} d x \right) \wedge d {\pi} .
\ee

From (\ref{domega}), we have
\be 
d u-D_{\pi} {\cal H} d x=0, \quad D_x {\pi}=-D_u {\cal H}.
\ee

As the result, we obtain 
\be \label{DWHE}
\frac{\partial u}{\partial x^{\mu}}=D_{\pi^{\mu}} {\cal H} , 
\quad \frac{\partial {\pi}^{\mu}}{\partial x^{\mu}}=-D_u {\cal H}, 
\ee
which are the well-known De Donder-Weyl Hamilton's equations (\ref{155}).

\subsection{Fractional Hamilton's equations}

The fractional generalization of the form (\ref{CP}) can be defined \cite{FDF,FV} by
\be \label{fCP}
\omega (\alpha)={\pi} d^{\alpha}_s u-{\cal H}(u,\pi) d^{\alpha}_s x .
\ee
It will be called the fractional Poincare-Cartan 1-form or simply 
the fractional action 1-form.
We can consider the fractional exterior derivative of the form (\ref{fCP}),
and use 
\be \label{domega-a}
d^{\alpha} \omega (\alpha)=0 
\ee
to obtain the fractional field equations. 
Here the fractional exterior derivative is
\be \label{d-a}
d^{\alpha}=d^{\alpha}_s x^{\nu} \, D^{\alpha}_{x^{\nu}}+d^{\alpha}_s u \, D^{\alpha}_u+
d^{\alpha}_s \pi^{\nu} \, D^{\alpha}_{\pi^{\nu}} , \quad
d^{\alpha}_s x= \Gamma^{-1}(\alpha+1) d^{\alpha} x^{\alpha} ,
\ee
where $D^{\alpha}_{x^{\nu}}$, $D^{\alpha}_u$, $D^{\alpha}_{\pi^{\nu}}$
can be fractional derivatives of different types \cite{KST}. 
For example, for $x^{\mu} \in \mathbb{R}^{2}$, such that $x=(t,r)$, we use
\[ D^{\alpha}_x=( \ _{t_0}^CD^{\alpha_0}_{t}, D^{\alpha_1}_r ) , \]
where $_{t_0}^CD^{\alpha_0}_{t}$ is Caputo fractional derivative, 
and $D^{\alpha_1}_r$ is the Riesz derivative. 
Fractional differential forms have been suggested in \cite{FDF}
and it is used to describe dynamical systems \cite{FV,JPA2005}. 

Then, by some transformations (see Appendix), one can obtain
\be \label{dfCP3}
d^{\alpha}\omega (\alpha)=\left[ D^{\alpha}_x {\pi}+D^{\alpha}_u {\cal H} \right]  
d^{\alpha}_s  x \wedge d^{\alpha}_s u- 
\left[ \frac{{\pi}^{1-\alpha}}{\Gamma(2-\alpha)} d^{\alpha}_s  u-
D^{\alpha}_{\pi} {\cal H} d^{\alpha}_s x \right] \wedge d^{\alpha}_s {\pi} .
\ee
Using (\ref{dfCP3}) and (\ref{domega-a}), we get 
\be \label{h-a}
\frac{{\pi}^{1-\alpha}}{\Gamma(2-\alpha)} d^{\alpha}_s u-
D^{\alpha}_{\pi} {\cal H} d^{\alpha}_s x=0, \quad 
D^{\alpha}_x {\pi}=-D^{\alpha}_u {\cal H} .
\ee
For the case $u=u(x)$, 
\be
d^{\alpha}_s u=d^{\alpha}_s x \, D^{\alpha}_x u .
\ee
As the result, we obtain
\be \label{fHe}
D^{\alpha}_{x^{\mu}} u= (\pi^{\mu})^{\alpha-1} \Gamma(2-\alpha) D^{\alpha}_{\pi^{\mu}} {\cal H} ,\quad
D^{\alpha}_{x^{\mu}} {\pi}^{\mu}=-D^{\alpha}_u {\cal H} .
\ee
These equations are the fractional generalization
of De Donder-Weyl Hamilton's equations (\ref{DWHE}).  

As an example, we can consider
\be
{\cal H}(u,\pi)=\frac{2-\alpha}{2} (\pi^0)^2-\frac{2-\alpha}{2} (\pi^1)^2+U(u) . 
\ee
Using 
\be
D^{\alpha}_{\pi^{\nu}} (\pi^{\mu})^{\beta}=
\frac{\Gamma(\beta+1)}{\Gamma(\beta+1-\alpha)} (\pi^{\mu})^{\beta-\alpha} \delta^{\mu}_{\nu},
\ee
and $\Gamma(z+1)=z\Gamma(z)$, equations (\ref{fHe}) give
\be \label{fHe2}
D^{\alpha_0}_t u= \pi^0 , \quad
D^{\alpha_1}_r u= -\pi^1 , 
\ee
\be \label{fHe3}
D^{\alpha_0}_t \pi_0+D^{\alpha_1}_t \pi_1=-D^{\alpha_u}_u U(u)  .
\ee
After substitution of (\ref{fHe2}) into (\ref{fHe3}), we obtain
\be \label{D2D2}
(D^{\alpha_0}_t)^2 u(t,r) -(D^{\alpha_1}_r)^2 u(t,r) +D^{\alpha_u}_u U(u)=0 .
\ee
For $\alpha_0=\alpha_1=\alpha_u=1$, Eq. (\ref{D2D2}) is the usual wave equation
\[ \partial^2_t u(t,r)-\partial^2_r u(t,r)+ \frac{\partial U(u)}{\partial u(t,r)}=0 . \]

For $U(u)=0$, Eq. (\ref{D2D2}) has the form
\be 
(D^{\alpha_0}_t)^2 u(t,r) -(D^{\alpha_1}_r)^2 u(t,r) =0 .
\ee
The solution  $u(t,r)$ of this equation is a linear combination of 
the solutions $u_1(t,r)$ and $u_2(t,r)$ of the equations
\be
D^{\alpha_0}_t u_1(t,r) -D^{\alpha_1}_r u_1(t,r) =0 , \quad
D^{\alpha_0}_t u_2(t,r) +D^{\alpha_1}_r u_2(t,r) =0 .
\ee
For $\alpha_1=1$, there exists a relation between the Dirac solutions and 
the fractional extension of D'Alembert expression that 
is considered in \cite{PV}.

Using the property
\be
D^{\alpha}_x D^{\alpha}_x u= D^{2\alpha}_x u ,
\ee
and we get the fractional equation
\be \label{ex1}
D^{2\alpha_0}_t u-D^{2\alpha_1}_r u+D^{\alpha_u}_u U(u)=0 .
\ee
For a special case,
\be
U(u)=m^2 \frac{\Gamma(2)}{\Gamma(2+\alpha_u)} u^{1+\alpha_u} ,
\ee
equation (\ref{ex1}) gives
\be \label{FKG}
D^{2\alpha_0}_t u(t,r)-D^{2\alpha_1}_r u(t,r)+ m^2 \, u(t,r)=0 .
\ee
This is a fractional generalization of Klein-Gordon equation 
(see also \cite{LZ,MBal}). 
Note that Eq. (\ref{FKG}) is not Lorentz invariant equations.
To obtain fractional relativistic equations, the fractional power of 
D'Alembertian should be used \cite{BG2} (see also Sec. 28 of \cite{SKM}).
The causality principle \cite{BS,Kempfle} also should be taken into account.

For 
\be
U(u)= \sin (u(t,r))
\ee
equation (\ref{ex1}) gives
\be
D^{2\alpha_0}_t u-D^{2\alpha_1}_r u+ \sin(u+\alpha_u \pi /2)=0 .
\ee
This equation is fractional sine-Gordon equation 
that was considered in \cite{LZ} for $\alpha_0=1$ and $\alpha_u=2$.

\section{Conclusion}

Exploiting a variational principle to obtain fractional dynamics
seems to be a fairly powerful tool that permits a universal consideration
of situations with fractal time and space.
In this paper, we have demonstrate it by deriving different equations
of using generalized Noether's theorem.
Such equations are similar to the regular conservation laws,
although the presence of fractional time derivatives reflects a dissipation.
Similar variation of action can be used to derive Hamiltonian type equations
although the situation here is not uniquely defined and some freedom
of choosing the differential action 1-form leaves different possibilities.

\section*{Acknowledge}

This work was supported by the Office of Naval Research, 
Grant No. N00014-02-1-0056, and the NSF Grant No. DMS-0417800.

\section*{Appendix}

To prove the proposition (\ref{dfCP3}), we use the rule 
\[ D^{\alpha}_{x} (fg) =\sum^{\infty}_{s=0} 
\left(^{\alpha}_s \right) (D^{\alpha-s}_x f )
\frac{\partial^s g}{\partial x^s} ,\]
and the relation \cite{KST}
\[ \frac{\partial^s}{\partial x^s} 
\left[ d^{\alpha}_s x \right]=0 \quad (s \ge 1) , \]
for integer $s$, where
\[ \left(^{\alpha}_k \right)=
\frac{(-1)^{k-1} \alpha \Gamma(k-\alpha)}{\Gamma(1-\alpha) \Gamma(k+1)}. \]
For example, we have 
\[ d^{\alpha} \left[ A_\mu d^{\alpha}_s x^{\mu} \right]=
\sum^{\infty}_{s=0}
d^{\alpha}_s x^\nu \wedge 
\left(^{\alpha}_s \right)
(D^{\alpha-s}_{x^\nu} A_\mu ) 
\frac{\partial^s}{\partial (x^\nu)^s} d^{\alpha}_s x^\mu=\]
\[ =d^{\alpha}_s x^\nu \wedge d^{\alpha}_s x^\mu
\left(^{\alpha}_0 \right) (D^{\alpha}_{x^\nu} A_\mu)=
\left( D^{\alpha}_{x^\nu} A_\mu \right)
d^{\alpha}_s x^\nu \wedge d^{\alpha}_s x^\mu . \]
As a result, 
\[ d^{\alpha}\omega (\alpha)=
d^{\alpha}( \pi d^{\alpha}_su) - d^{\alpha}({\cal H}d^{\alpha}_sx)=\]
\[ =(D^{\alpha}_x {\pi}) d^{\alpha}_s x \wedge d^{\alpha}_s u+
(D^{\alpha}_u {\pi}) d^{\alpha}_s u \wedge d^{\alpha}_s u+
(D^{\alpha}_{\pi} {\pi}) d^{\alpha}_s {\pi} \wedge d^{\alpha}_s u- \]
\be \label{dfCP1} -
(D^{\alpha}_x {\cal H}) d^{\alpha}_s x \wedge d^{\alpha}_s x- 
(D^{\alpha}_u {\cal H}) d^{\alpha}_s u \wedge d^{\alpha}_s x-
(D^{\alpha}_{\pi} {\cal H}) d^{\alpha}_s\pi \wedge d^{\alpha}_s x .
\ee
Here $D^{\alpha}_{x^0}= \ _{t_0}^CD^{\beta}_{t}$ 
is Caputo fractional derivative,
and $D^{\alpha}_{x^1}=\partial^{\alpha} / \partial |r|^{\alpha}$ 
is the Riesz fractional derivative \cite{KST}.
Using 
\[ d^{\alpha}_s {\pi} \wedge d^{\alpha}_s x=-d^{\alpha}_s x \wedge d^{\alpha}_s {\pi} \]
and $D_x{\cal H}(u,\pi)=0$, $D^{\alpha}_u {\pi}=0$ for Riesz and Caputo derivatives, 
we can rewrite equation (\ref{dfCP1}) in the form
\be \label{dfCP2}
d^{\alpha} \omega (\alpha)=
\left[ D^{\alpha}_x {\pi}+ D^{\alpha}_u {\cal H} \right] 
d^{\alpha}_s x \wedge d^{\alpha}_s u-
\left[ (D^{\alpha}_{\pi} {\pi}) d^{\alpha}_s u- (D^{\alpha}_{\pi} {\cal H}) d^{\alpha}_s x 
\right] \wedge d^{\alpha}_s {\pi} .
\ee
Substitution of 
\be
D^{\alpha}_{\pi} {\pi}=\frac{{\pi}^{1-\alpha}}{\Gamma(2-\alpha)},
\ee
into equation (\ref{dfCP2}) gives (\ref{dfCP3}).


\end{document}